# Design of spontaneous parametric down-conversion in integrated hybrid Si$_x$N$_y$-PPLN waveguides

XIANG CHENG[1,2,4†], MURAT CAN SARIHAN[2†], KAI-CHI CHANG[2†], YOO SEUNG LEE[2,5], FABIAN LAUDENBACH[3], HAN YE[1], ZHONGYUAN YU[1,6], AND CHEE WEI WONG[2]

[1]*State Key Laboratory of Information Photonics and Optical Communications, Beijing University of Posts and Telecommunications, P.O. Box 72, Beijing 100876, PR China*
[2]*Fang Lu Mesoscopic Optics and Quantum Electronics Laboratory, Department of Electrical Engineering, University of California, Los Angeles, CA 90095, USA*
[3]*Security & Communication Technologies, Center for Digital Safety & Security, AIT Austrian Institute of Technology GmbH, Giefinggasse 4, 1210 Vienna, Austria*
[4]*e-mail: cxmamba@outlook.com*
[5]*e-mail: yooseunglee@ucla.edu*
[6]*e-mail: yuzhongyuan30@hotmail.com*
[†]*These authors contributed equally.*

**Abstract:** High-efficient and high-purity photon sources are highly desired for quantum information processing. We report the design of a chip-scale hybrid Si$_x$N$_y$ and thin film periodically-poled lithium niobate waveguide for generating high-purity type-II spontaneous parametric down conversion (SPDC) photons in telecommunication band. The modeled second harmonic generation efficiency of 225% W$^{-1}$ • cm$^{-2}$ is obtained at 1560nm. Joint spectral analysis is performed to estimate the frequency correlation of SPDC photons, yielding intrinsic purity with up to 95.17%. The generation rate of these high-purity photon pairs is estimated to be 2.87 × 10$^7$ pairs/s/mW within the bandwidth of SPDC. Our chip-scale hybrid waveguide design has the potential for large scale on-chip quantum information processing and integrated photon-efficient quantum key distribution through high-dimensional time-energy encoding.

## 1. Introduction

Rapid development in quantum information technology demands bright, high-efficient and integrated quantum light source, which scales and compacts the optics system [1-5]. By taking advantage of large $\chi^{(2)}$ coefficient and broad transparent window (350 nm to 4.5 μm) of lithium niobate (LiNbO$_3$, LN) [6], spontaneous parametric down conversion (SPDC) has been demonstrated to be a good candidate for non-classical light generation in the telecommunication band. Conventional SPDC devices are typically realized by titanium-indiffused periodically-poled lithium niobate (PPLN) waveguides, where quasi-phase matching (QPM) condition is achieved by periodic domain inversion [7-9]. However, these devices suffer from weak optical modes confinement due to the low index contrast between waveguide core and cladding, which also results in large device dimensions. In addition, the waveguide modes are different at fundamental wavelength and SPDC wavelength because of asymmetric gradient-index profile [8]. Thus, the poor overlap between waveguide modes significantly reduces the SPDC efficiency. On the other hand, recent progress in nonlinear photonics provides methods that enhance nonlinear interaction by orders of magnitude due to the remarkable light confinement in these wavelength-scale devices [10-12]. State-of-the-art nanofabrication technology makes it possible to construct scalable, integrated and economical nonlinear optical system [13].

Conventional photonic circuits based on Si waveguides have been investigated and developed thanks to broadly established CMOS fabrication techniques [14,15]. However, Si suffers from a high nonlinear absorption due to two-photon absorption (TPA) in telecommunication bands with wavelength lower than ~2 μm. Also, Si's intrinsic nonlinear figure of merit ($NFOM = n_2/(\beta \times \lambda)$, where $n_2$ is the Kerr coefficient, $\beta$ is the TPA coefficient and $\lambda$ is the wavelength) near 1550nm is only ~0.3 [16]. These fundamental limitations result from the intrinsic property of Si' band structure. On the other hand, recently developed platforms based on silicon nitride ($Si_3N_4$) have made significant progress with respect to nonlinear performance. $Si_3N_4$, a CMOS-compatible material with no TPA, shows low propagation loss for telecommunication wavelengths and high third-order nonlinearity, which can be a very promising nonlinear platform for integrated photonic circuits [17].

In this work, we design a hybrid QPM waveguide structure of Si-rich silicon nitride ($Si_xN_y$) waveguide and thin film PPLN for generating high-purity type-II SPDC photons, which combines two different device platforms and enables multifunctional chip-scale and wafer-level integration. For normal $Si_3N_4$, optical modes of the hybrid waveguide are mostly confined inside LN film because the refractive index of $Si_3N_4$ is smaller than that of the LN [13,18]. By using $Si_xN_y$ (with higher Si/N ratio than $Si_3N_4$), optical mode can be transferred from LN film to $Si_xN_y$ waveguide so that the reflection loss can be reduced at between the waveguide and the LN film. Compared with other waveguide proposals [13, 19-21], the LN thin film can be deposited on a large scale photonic circuit and periodically poled to realize SPDC photon source without off-chip coupling due to the mode transfer design. By tuning the waveguide width and optimizing refractive index of $Si_xN_y$ [22, 23], optical modes can be tightly confined in PPLN region of designed hybrid waveguide structure. These advantages can significantly enhance the nonlinear interaction thus improve the efficiency of nonlinear conversion, and modeled second harmonic generation (SHG) efficiency of 225% $W^{-1} \cdot cm^{-2}$ is obtained at 1560 nm. The phase matching condition is then optimized by tuning the pump and the length of PPLN to achieve high purity biphoton state. Joint spectral analysis of the SPDC photon pairs is performed to estimate spectral purity, and we yield a high-purity of 95.17% without band-pass filtering. Based on the optimal waveguide design, we estimate a photon pair generation rate of 2.87 × $10^7$ pairs/s/mW within the bandwidth of SPDC. Our hybrid waveguide design can serve as high-purity type-II SPDC photon source for integrated quantum information processing and quantum communications.

## 2. Waveguide structure design

The structure of our designed hybrid Si-rich silicon nitride and thin film PPLN waveguide is illustrated in Fig. 1(a). We specially design the hybrid waveguide for type-II SPDC to generate photon pairs with orthogonal polarization that can be separated efficiently by a polarization beamsplitter for convenient experimental implementation. This hybrid waveguide device can be fabricated based on LNOI (LN on isolator) substrate wafer which is already commercially available [24]. $Si_xN_y$ waveguide with modified Si/N ratio can be manufactured on a $SiO_2$-on-Si substrate wafer via plasma enhanced chemical vapor deposition or low-pressure chemical vapor deposition process, which is CMOS-compatible [22, 23]. The cross-section of the hybrid waveguide device is schematized in Fig. 1(b). The $Si_xN_y$ waveguide is 800-nm thick with a 700-nm thick LN thin film deposited above. The LN thin film is *x*-cut oriented so that TE mode field of the waveguide is along the extraordinary axis. The hybrid waveguide device consists of five sections from input to output. Section A is the input waveguide region. The width of the $Si_xN_y$ waveguide is optimized as 1.2 μm to guide the fundamental TM mode of 780nm pump photons with lower loss. Section B is the input mode transition region. The $Si_xN_y$ waveguide in this section is designed as tapered waveguide with width decreasing from 1.2 μm to 400 nm. Along the 200-μm long tapered waveguide, the optical modes will be lifted and gradually become confined in the LN layer for higher SPDC efficiency. Section C is the PPLN region for SPDC photon generation. In this section, the optical modes are tightly confined in LN thin film to enhance the nonlinear interaction by optimizing the waveguide width and refractive index of $Si_xN_y$. Mode confinement and mode overlap are investigated in this active region to characterize the conversion efficiency. Section D is the output mode transition region. The Si$_x$N$_y$ waveguide in this section is a 200- μm long tapered waveguide with width increasing from 400 nm to 2 μm. The optical modes fall into Si$_x$N$_y$ waveguide for convenient output. Section E is the output waveguide region with Si$_x$N$_y$ waveguide width optimized as 2 μm for efficiently guiding the fundamental modes of the SPDC photons.

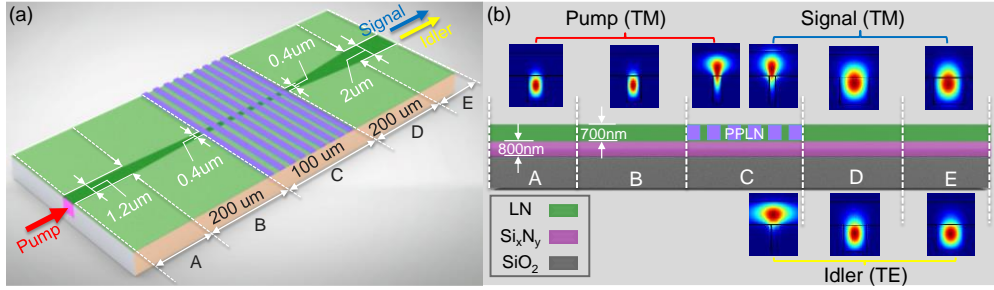

Fig. 1. (a) Illustration of chip-scale hybrid Si-rich silicon nitride and thin film PPLN waveguide structure for type-II SPDC (780 nm (o) → 1560 nm (o) + 1560 nm (e)). The hybrid waveguide consists of five sections, labelled from A to E. Dimensions of the hybrid waveguide structure are denoted for each section. (b) Cross-section view of hybrid waveguide structure. Mode profiles in each section of the device are simulated for the pump and SPDC photons. Opical mode is lifted into PPN region from section A to section C, and goes down to Si$_x$N$_y$ waveguide from section C to section E.

The simulated fundamental mode profiles of the hybrid waveguide device for pump photons (780 nm) and SPDC photons (1560 nm) in different sections are depicted in Fig. 1(b). We can observe from the mode profiles that the optical mode of pump is first confined into the Si$_x$N$_y$ waveguide, and then lifted to PPLN region in section C. The SPDC light generated from the PPLN region goes down to the Si$_x$N$_y$ waveguide for convenient output with over 92.3% transmission. Thanks to the high index contrast structure and submicrometer thickness of the LN thin film, the waveguide modes are tightly confined into a mode area that is smaller than that achieved with proton exchange or Ti indiffusion [25, 26]. This mode confinement feature directly relates to the refractive index of Si$_x$N$_y$. Table 1 shows the refractive indices of four different Si$_x$N$_y$ materials for 780 nm and 1560 nm, and LN for $o$ light (TM) or $e$ light (TE), along with the normal silicon nitride for comparison. The refractive index profiles of these four Si$_x$N$_y$ materials are approximated according to previous ellipsometry studies [22, 27] by

$$\frac{N}{Si} = \frac{4}{3} \frac{n_{Si} - n}{n + n_{Si} - 2n_{Si_3N_4}}, \qquad (1)$$

where $n$, $n_{Si}$ and $n_{Si_3N_4}$ represent the refractive indices of Si$_x$N$_y$, Si and Si$_3$N$_4$, respectively. Four Si$_x$N$_y$ materials are assigned with increasing Si/N ratio, which leads to increasing refractive index from Material 1 to Material 4.

Table 1. Refractive indices of Si$_x$N$_y$ with different stoichiometric ratio.

| Material | n | | Si/N ratio |
|---|---|---|---|
| | 780 nm | 1560nm | |
| Mat. 1 (Si$_x$N$_y$) | 2.225 | 2.174 | 0.9556 |
| Mat. 2 (Si$_x$N$_y$) | 2.260 | 2.205 | 0.9975 |
| Mat. 3 (Si$_x$N$_y$) | 2.282 | 2.225 | 1.025 |
| Mat. 4 (Si$_x$N$_y$) | 2.325 | 2.266 | 1.081 |
| Si$_3$N$_4$ | 2.025 | 1.996 | 0.75 |
| LN($e$) | 2.178 | 2.137 | |
| LN($o$) | 2.258 | 2.211 | |

## 3. Mode analysis and conversion efficiency

In order to investigate the impact of different $Si_xN_y$ materials on the conversion efficiency, we first analyze the mode area and mode confinement of each constituent mode of SPDC process, then calculate the mode overlap area ($S_{eff}$) of pump and SPDC light and estimate the second-harmonic generation normalized efficiency of the hybrid waveguide device for four materials. The effective refractive indices for pump and SPDC light are computed via eigenmode analysis for the waveguide structure. The mode profiles are simulated by Finite-Difference Time-Domain method in detail for each eigenmode, and used to analyze mode area, confinement factor and mode overlap.

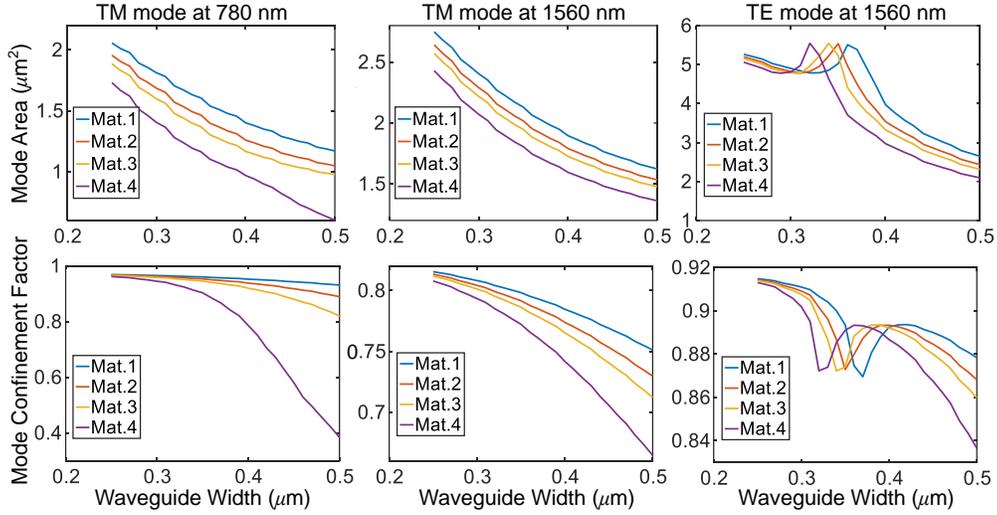

Fig. 2. Mode area and confinement characteristics of each constituent mode of type-II SPDC for different $Si_xN_y$ materials as functions of waveguide width. The thickness of the waveguide is fixed to 800 nm.

The mode area $A_{mode}$ is calculated for each individual mode via the following equation:

$$A_{mode} = \frac{\left(\iint E^2(x,y)dxdy\right)^2}{\iint |E^2|^2(x,y)dxdy}, \quad (2)$$

where E(x,y) is the spatial mode profile of transverse waveguide modes. Mode confinement factor is calculated to evaluate the fraction of power confined into LN region which contributes to SPDC generation. The mode confinement factor is given by:

$$\Gamma = \frac{\iint d(x,y)E^2(x,y)dxdy}{\iint E^2(x,y)dxdy}, \quad (3)$$

here, d(x,y) represents the normalized nonlinear coefficient. The mode area and confinement factor of pump and SPDC light are analyzed for different $Si_xN_y$ material in Fig. 2. Here, it can be seen that there is a tradeoff between low mode area and strong confinement. When the Si/N ratio in the $Si_xN_y$ material increases, mode area becomes smaller hence more confined to the nitride ridge due to higher refractive index contrast. The effect of this mode shrinkage is a power draw from the active region that will significantly harm the nonlinear interaction. Thus, it is important to optimize the structure to get a balanced mode size and confinement. We note that there are abrupt peaks observed for the mode area of TE mode (idler) for different $Si_xN_y$ materials. The reason is that the idler mode fields are not purely TE and have $E_y$ fraction close to 50% purity, while the TM (signal) modes still preserve its purity at 1560 nm for the same waveguide configurations. This quasi-TE nature of idler modes leads to a less concentrated mode field profile, and results in larger mode area and weaker confinement. $Si_xN_y$ has advantages over normal $Si_3N_4$ since the refractive index can be engineered to be larger than the index of LN, while preserving the compatibility. If the index contrast is optimized, the mode confinement can then be tuned by adjusting the $Si_xN_y$ waveguide width throughout the chip.

Next, we examine the mode overlap between the pump and SPDC light for different $Si_xN_y$ materials, and then calculate the normalized efficiency. Mode overlap area integral ($S_{eff}$) between waveguide modes at ω (signal and idler) and 2ω (pump) is defined as [13]:

$$S_{eff} = \frac{\left[\iint E_p^2(x,y)dxdy\right]\left[\iint E_s^2(x,y)dxdy\right]\left[\iint E_i^2(x,y)dxdy\right]}{\left[\iint d(x,y)E_p(x,y)E_s(x,y)E_i(x,y)dxdy\right]^2}. \quad (4)$$

This coefficient confines the calculation of denominator to the active region (PPLN), while the numerator is calculated over the whole region. The SHG normalized efficiency $\eta_{nor}$ is expressed as:

$$\eta_{nor} = \frac{8d_{eff}^2}{\varepsilon_0 c n_p n_s n_i \lambda_p^2 S_{eff}} \mathrm{sinc}^2(\frac{\Delta k L}{2}), \quad (5)$$

where Δk is the phase mismatch, L is the length of nonlinear region, $\varepsilon_0$ is the vacuum permittivity, c is the speed of light in vacuum, $n_p$, $n_s$, and $n_i$ are the effective refractive indices for pump (p), signal (s) and idler (i), respectively [13]. The parameter $d_{eff}$ is the nonlinear coefficient of the LN, which is changing anisotropically according to the type of quasi-phase matching. In this study, we use $d_{eff} = 2d_{31}/\pi = 2.8$ pm/V for type-II QPM condition [28]. Small $S_{eff}$ means that the pump and signal/idler modes are well-confined and overlapping each other at PPLN region. For our hybrid waveguide, the material and structural properties are designed to ensure small mode fields for minimum mode overlap area to get higher conversion efficiency thus brightness.

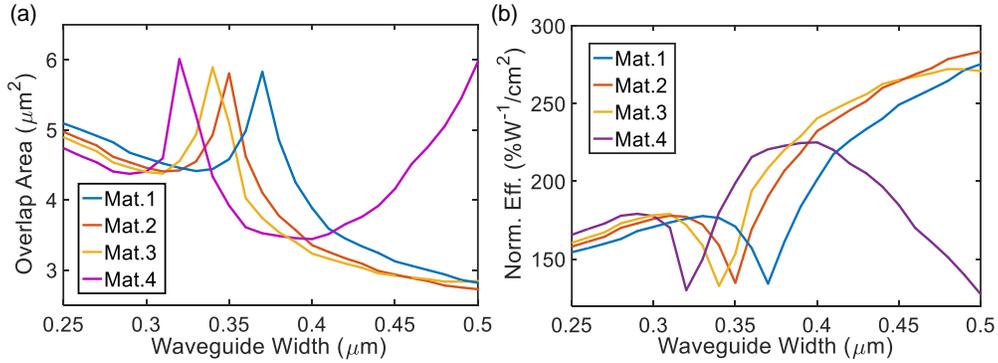

Fig. 3. (a) Mode overlap area between fundamental modes of pump and SPDC photons for different $Si_xN_y$ materials and varying waveguide width. (b) Second-harmonic generation normalized efficiency of the hybrid waveguide for different $Si_xN_y$ materials and varying waveguide width.

Lower $S_{eff}$ indicates that the field diameters for each optical mode are small and the centers of each mode at z-axis are close to each other. It is also important to realize the mode transfer between $Si_xN_y$ waveguide and LN thin film, which can enable dispersion engineering and benefit chip-scale integration. The peak normalized efficiency can be obtained when the optical mode is confined to PPLN region under this consideration. Different materials with increasing Si/N ratio (higher refractive index) are examined for this purpose. The mode overlap area $S_{eff}$ between pump, signal and idler photons for different $Si_xN_y$ materials are depicted in Fig. 3(a). When the $Si_xN_y$ waveguide thickness is 800 nm, we observed that Material 4 with highest Si/N ratio has the lowest $S_{eff}$ until enlarging the waveguide width up to around 400 nm. Then, due to the nonlinear characteristics, Material 2 and 3 has lower overlap area than Material 4. For larger waveguide widths, Material 2 and 3 are converging to each other in terms of overlap area. We also observe spikes for four different $Si_xN_y$ materials in Fig. 3(a), resulting from the similar reason as in Fig. 2. The quasi-TE nature of idler modes enlarges the mode area thus leads to larger mode overlap.

The SHG normalized efficiencies are calculated based on the mode overlap area and individual mode indices, and shown in Fig. 3(b). Here, the trend is similar as Fig. 3(a). But due to very low overlap area and lower mode indices, which is determined by index properties of comprising materials for the hybrid structure, Material 2 stands out with higher normalized efficiency for wider waveguides. Also according to the trend, it is expected that Material 1 has the best properties since it has the lowest bulk refractive index due to lower Si/N ratio. However, the case for Material 1 is special because the refractive index of Material 1 is in between the indices of LN for *e* and *o* light with a good margin. The pump mode for Material 1 is less confined into the $Si_xN_y$ waveguide and spread more into the LN layer, which introduces more loss thus lower efficiency than that of Material 2. Furthermore, the anisotropic behavior shows itself in the Fig. 3(a) as a sharp divergence between 300 - 400 nm waveguide width from the trend of TE modes at extraordinary axis. The reason is that for this waveguide configuration, the optical mode of idler photons at 1560 nm is quasi-TE mode with a lower $E_y$ fraction, which result in similar trend in Fig. 2. Overall, Material 4 has the advantage that optical modes can be engineered to go up and down along propagation between $Si_xN_y$ waveguide and LN thin film, which benefits large scale on-chip integration. The optical mode can be lifted to active region for efficient SPDC photon generation, and the SPDC light can propagate through $Si_xN_y$ waveguides over the photonic circuit. If we only consider conversion efficiency, Material 2 offers a higher normalized efficiency but the mode area in LN region will decrease, and optical modes will exhibit similar propagation trend as in ridge waveguide structure [13,18].

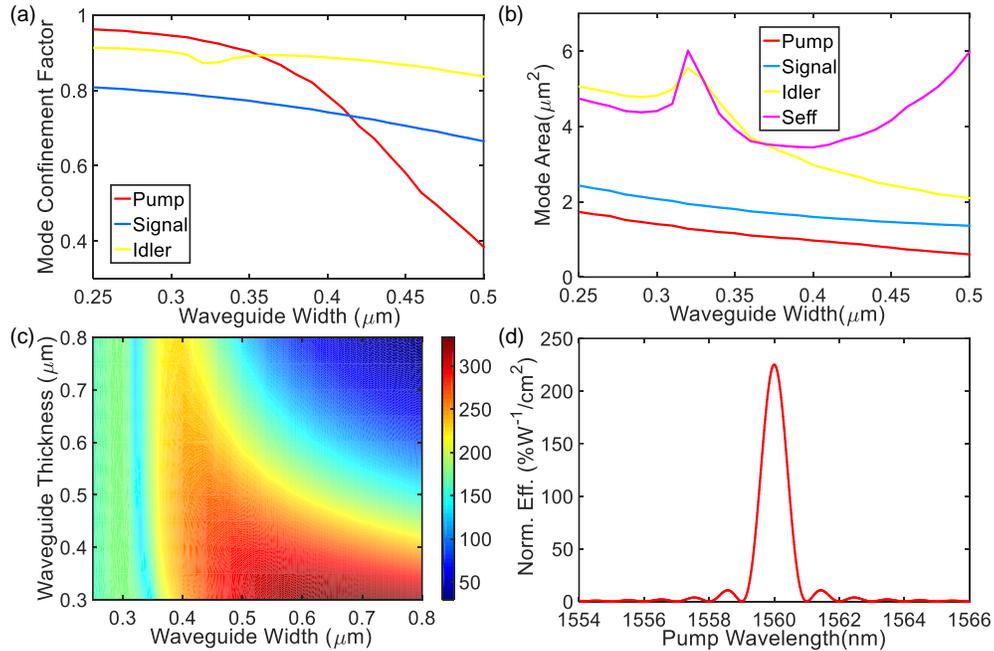

Fig. 4. For $Si_xN_y$ material 4: (a) Mode confinement factor of pump, signal and idler photons as functions of waveguide width. (b) Mode area of pump and SPDC photons and mode overlap area with varying waveguide width. (c) SHG normalized efficiency as a function of $Si_xN_y$ waveguide width and thickness. (d) SHG normalized efficiency as a function of the pump wavelength, with 400 nm waveguide width and 800 nm waveguide thickness.

Then, we analyze the mode properties individually for Material 4, and estimate confinement into active region, mode area and normalized efficiency in Fig. 4. Furthermore, a complete design map for different waveguide width and thickness is provided. Fig. 4(a) shows the variance in power confinement to the PPLN region. Here, the confinement decreases with increasing waveguide width while mode area and $S_{eff}$ gets smaller. Fig. 4(b) indicates that mode overlap area $S_{eff}$ is dominated by the mode with largest area, which is idler (TE). The TM modes lying on the ordinary axis decrease monotonically, while idler shows the abovementioned trend due to anisotropy as in Fig. 2. It is noted that $S_{eff}$ increases after 400 nm waveguide width while idler mode becomes smaller. The reason is that the optical mode becomes more confined to $Si_xN_y$ waveguide, thus affects the overlap in the LN region. Fig. 4(c) shows the design map that includes the effect of $Si_xN_y$ waveguide thickness on the normalized efficiency. Increasing waveguide thickness diminishes the mode fraction confined into LN thin film, thus it needs to be optimized to achieve better confinement while keeping mode areas as small as possible. The highest efficiency is obtained for the optimal waveguide parameters of 800 nm thickness and 400 nm width with a peak efficiency of 225% $W^{-1} \cdot cm^{-2}$. It is also possible to obtain a peak efficiency of 333.6% $W^{-1} \cdot cm^{-2}$ by lowering the waveguide width to 300 nm with Material 4, but the abovementioned mode lifting behavior for chip-scale integration no longer exists. The spectral dependence of the SHG normalized efficiency with the optimal waveguide structure is also depicted in Fig. 4(d). The Full-Width-Half-Maximum is ~1 nm. The normalized efficiency is proportional to a cardinal sine squared function, assuming the waveguide is uniform [13].

## 4. Joint spectral analysis

Next, we verify the frequency correlation of the SPDC photons generated from hybrid waveguide by performing joint spectral analysis, which reveals the degree of correlation between signal and idler photons [29-31]. In the SPDC process, one pump photon at frequency $\omega_p$ is annihilated in the medium and two photons (signal photon at $\omega_s$, idler photon at $\omega_i$) are spontaneously created. The quantum state that describes SPDC can be expressed as [28]:

$$|\Psi\rangle = \tilde{N} d_{eff} L \int_0^\infty \int_0^\infty \varphi(\omega_s + \omega_i) \phi(\omega_s + \omega_i) a_s^\dagger a_s^\dagger d\omega_i d\omega_s |0\rangle, \quad (6)$$

where $\tilde{N}$ is a normalization constant, $d_{eff}$ is the effective nonlinear coefficient of the crystal, L is the length of the crystal, $a_s^\dagger$ and $a_i^\dagger$ are the creation operators of signal and idler photon respectively. $\varphi$ is the pump envelope amplitude, and $\phi$ represents the phase-matching envelope amplitude as:

$$\phi(\omega_s, \omega_i) = e^{\frac{i\Delta kL}{2}} \text{sinc}(\frac{\Delta kL}{2}), \quad (7)$$

with $\Delta k$ representing the phase mismatch:

$$\Delta \hat{k} = 2\pi(\frac{n_p}{\lambda_p}\hat{k}_p - \frac{n_s}{\lambda_s}\hat{k}_s - \frac{n_i}{\lambda_i}\hat{k}_i), \quad (8)$$

where $\lambda_p / \lambda_s / \lambda_i$ and $n_p / n_s / n_i$ are the wavelengths and refractive indices for pump, signal and idler, respectively. For collinear quasi-phase-matching, the mismatch can be written as:

$$\Delta k = 2\pi(\frac{n_p}{\lambda_p} - \frac{n_s}{\lambda_s} - \frac{n_i}{\lambda_i} - \frac{m}{\Lambda}), \quad (9)$$

where m is an odd integer referred to as QPM order, $\Lambda$ is the poling periodicity of the PPLN. The refractive indices are obtained according to the wavelength- and temperature- dependent Sellmeier equations at room temperature.

The joint spectral amplitude (JSA) is defined as the product of the pump envelope amplitude $\varphi$, and the phase-matching envelope amplitude $\phi$ [28, 29]:

$$f(\omega_s, \omega_i) = \varphi(\omega_s + \omega_i)\phi(\omega_s, \omega_i). \quad (10)$$

The joint spectral intensity (JSI) can then be defined as $\text{JSI}=|\text{JSA}|^2$, which can be easily carried out in experiment by utilizing fiber dispersion to measure the frequency correlation of SPDC photon pairs based on their arrival time [32]. The amplitude of SPDC can thus be rewritten as:

$$|\Psi\rangle = \tilde{N} d_{eff} L \int_0^\infty \int_0^\infty f(\omega_s, \omega_i) a_s^\dagger a_i^\dagger d\omega_s d\omega_i |0\rangle. \quad (11)$$

To study the degree of entanglement of $|\Psi\rangle$, we first express the SPDC photons using reduced density matrix:

$$\hat{\rho}_s = \text{Tr}_i(\hat{\rho}), \qquad \hat{\rho}_i = \text{Tr}_s(\hat{\rho}), \quad (12)$$

where $\text{Tr}_s$ ($\text{Tr}_i$) is the partial trace over the subsystem of signal (idler), and $\hat{\rho} = |\Psi\rangle\langle\Psi|$. The purity of the SPDC photons is defined as:

$$P_s = \text{Tr}(\hat{\rho}_s^2), \qquad P_i = \text{Tr}(\hat{\rho}_i^2). \quad (13)$$

In order to reveal the influence of spectral correlation in the pure bipartite state $|\Psi\rangle$ on the purity, we estimate the Schmidt decomposition of the joint two-photon state. The Schmidt decomposition can be numerically computed by the singular value decomposition [28, 29]. We express the Schmidt decomposition in terms of a complete set of basis states as:

$$|\Psi\rangle = \sum_j \sqrt{\lambda_j} |s_j\rangle \otimes |i_j\rangle, \qquad \sum_j \lambda_j = 1. \quad (14)$$

The orthonormal basis states $|s_j\rangle$ and $|i_j\rangle$ are the Schmidt modes representing the respective subsystem of signal and idler. $\lambda_j$ is the Schmidt coefficient for each Schmidt mode. The Schmidt decomposition provides an intuitive measurement of entanglement of a pure state. This can be quantified by the Schmidt number, K, which is defined as:

$$K = \frac{1}{\sum_j \lambda_j^2} = \frac{1}{\text{Tr}(\hat{\rho}_s^2)} = \frac{1}{\text{Tr}(\hat{\rho}_i^2)}. \quad (15)$$

If a product state is entangled, there is more than just one Schmidt mode and K > 1. A product state would be unentangled and can be expressed with only one Schmidt mode, thus S equals unity. A maximally entangled product state requires an infinite number of Schmidt modes to describe so that S equals infinity.

According to Schmidt number and purity expression, the purity of the reduced photon states can thus be calculated by the inverse of the Schmidt number:

$$P_s = P_i = \sum_j \lambda_j^2 = \frac{1}{K}. \quad (16)$$

The purity of the photon states is proportional to the reciprocal of S. In order to obtain pure photon states, there should be only one non-vanishing Schmidt coefficient $\lambda = 1$ yielding maximum purity $P_s = P_i = 1$. In this case, the signal and idler quantum states are pure and the SPDC amplitude can be separately expressed for $\omega_s$ and $\omega_i$ as:

$$|\Psi\rangle = |s\rangle \otimes |i\rangle = \tilde{N} d_{eff} L \int_0^\infty f_s(\omega_s) a_s^\dagger d\omega_s \int_0^\infty f_i(\omega_i) a_i^\dagger d\omega_i |0\rangle. \quad (17)$$

In order to achieve high-purity of the SPDC photons, the signal and idler photons should be as uncorrelated as possible. The pump and the phase-matching envelope amplitude should also be matched to produce a frequency-indistinguishable JSI, thus yield high purity. This could be achieved by adjusting the pump pulse duration and the length of PPLN.

The design map of purity for different PPLN length and pump pulse duration is shown in Fig. 5(a). High purity can be obtained with short PPLN length and pulse duration. But there is a trade-off between high efficiency and high purity, since shorter length of PPLN will reduce the SPDC photons generated from our hybrid waveguide device. For $Si_xN_y$ waveguide width and thickness of 400 nm and 800 nm, we obtain the highest purity of 95.17% with 100 μm PPLN length and 64 fs pulse duration. The poling period is set to 4.3 μm to achieve phase-matching condition. This high purity indicates that the SPDC photons are highly uncorrelated intrinsically. Fig. 5(b) illustrates the JSI of these high-purity SPDC photons without band-pass filtering. The spectra of signal and idler photons are matched so that they are frequency-indistinguishable. This spectral indistinguishability of the SPDC photons is crucial to achieve high-visibility two-photon interference, especially between one photon pair's signal and other pair's idler [28,33].

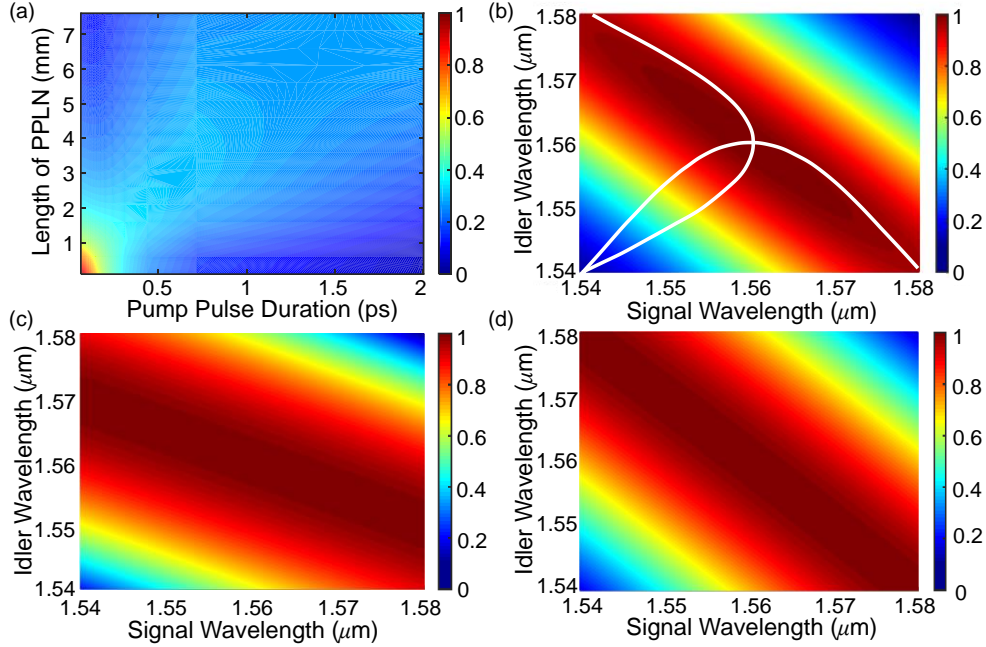

Fig. 5. (a) Purity of type-II SPDC photon pairs as a function of pump pulse duration and length of PPLN. (b) JSI distribution of SPDC photon pairs with 100 μm length of PPLN and 0.064 ps pump pulse duration, yielding purity of 95.17% without filtering. The spectra of signal and idler are depicted by white curves. (c) phase-matching envelope intensity and (d) pump envelope intensity for the composed JSI. The width and thickness of $Si_xN_y$ waveguide is fixed at 400 nm and 800nm, respectively.

The phase-matching envelope intensity $|\phi|^2$ and pump envelope intensity $|\varphi|^2$ for the JSI are also shown in Fig. 5(c) and 5(d). We notice that the direction of the pump envelope intensity is tilted at -45 ° (with positive direction of the horizontal axis), while the direction of the phase-matching envelope intensity can be at arbitrary angle θ. The -45 ° angle of the pump envelope intensity direction comes from the energy conservation of the SPDC process, whereas the arbitrary angle θ of the phase-matching envelope intensity direction can be determined by the group velocities as [33]:

$$\tan\theta = \frac{v_p^{-1} - v_s^{-1}}{v_p^{-1} - v_i^{-1}}, \tag{18}$$

where $v_p$, $v_s$, $v_i$ are the group velocities of the pump, signal and idler, respectively. With the waveguide parameter combination for high purity of 95.17%, the direction of phase-matching envelope intensity is calculated to be tilted at θ ≈ -13 ° as shown in Fig. 5(c). This tilted angle of phase-matching envelope intensity leads to the elliptical shape of the JSI [34, 35]. For comparison, near-circular shape of JSI is easier to achieve with periodically-poled potassium titanyl phosphate (PPKTP) waveguide due to diagonal phase-matching envelop amplitude [28, 36]. However, the intrinsic purity of the SPDC photons from PPKTP waveguide is limited to

~83% because of the sidelobes of the sinc function in joint spectral amplitude [36]. For our hybrid waveguide design, we manage to obtain high intrinsic purity despite of the elliptical shape of JSI by optimizing the phase-matching condition. Although the SPDC photons generated from hybrid waveguide have intrinsically high purity, a band-pass filter can still be implemented in order to achieve near-unity two-photon interference visibility in experiments.

Based on the current waveguide configuration, we can estimate the generated SPDC signal power by [37, 38]:

$$dP_s = \frac{16\pi^3 \hbar d_{eff}^2 L^2 c P_p}{\varepsilon_0 n_p n_s n_i \lambda_s^4 \lambda_i S_{eff}} \operatorname{sinc}^2\left(\frac{\Delta k L}{2}\right) d\lambda_s, \tag{19}$$

here, $P_p$ is the pump power. We use $d_{eff}$ = 2.8 pm/V, L = 100 μm, $n_p$ = 2.325, $n_s$ = $n_i$ = 2.266, $\lambda s$ = $\lambda s$ = 1560 nm, $P_p$ = 1 mW. For the designed waveguide of 800 nm thickness and 400 nm width, we have $S_{eff}$ = 3.441 μm$^2$ as calculated from the mode overlap and estimate a pair generation rate of $2.87 \times 10^7$ pairs/s/mW of pump power within the SPDC bandwidth.

## 5. Conclusion

In summary, we designed a chip-scale hybrid $Si_xN_y$ and thin film PPLN waveguide to generate high-purity type-II SPDC photons. By utilizing $Si_xN_y$, we achieved mode transition from $Si_xN_y$ waveguide to LN thin film, which enables multifunctional integrated photonic circuit and chip-scale quantum system. The hybrid waveguide device shows a peak normalized efficiency of 225% W$^{-1}$·cm$^{-2}$ at 1560nm for SHG. Joint spectral analysis of the SPDC photons provides an estimation of purity with up to 95.17% under the phase matching condition without band-pass filtering. The photon pair generation rate of our device is estimated to be $2.87 \times 10^7$ pairs/s/mW within the bandwidth of SPDC. Our high-efficiency chip-scale hybrid waveguide can work as an integrated SPDC photon source for on-chip quantum information processing and practical secure quantum key distribution channels.

## Appendix: High-dimensional biphoton frequency comb

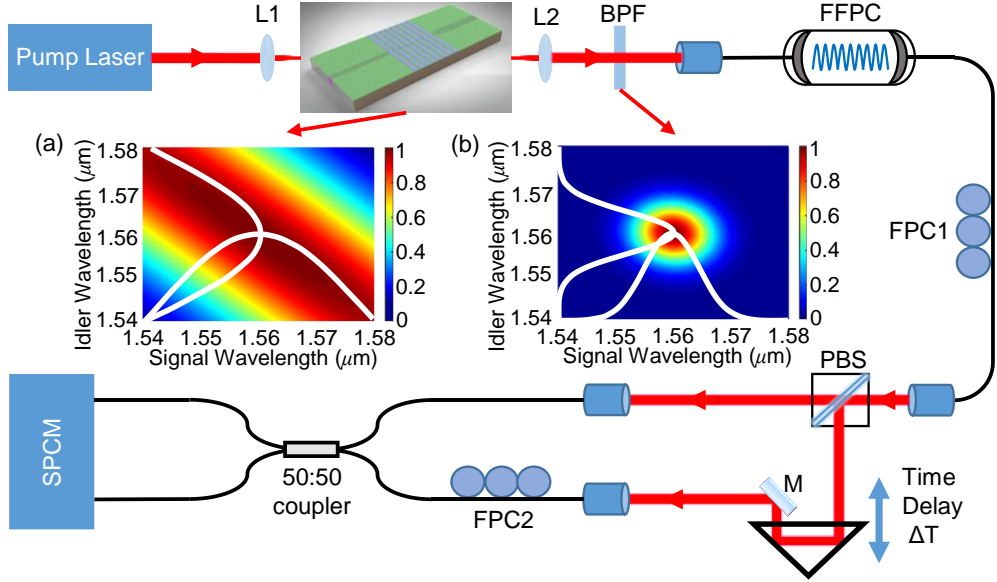

Fig. 6. Experimental scheme for generating high-dimensional biphoton frequency comb and observing HOM quantum revival. L: lens; BPF: band-pass filter; FFPC: fiber Fabry-Pérot cavity; FPC: fiber polarization controller; PBS: polarization beamsplitter; M: reflective mirror; SPCM: single photon counting module. Inset: (a) JSI distribution of the SPDC photons generated by our designed hybrid waveguide. (b) JSI distribution of the filtered photons, with purity up to 99.79%. The spectra of the signal and idler photons are indicated by the white curves.

Here, we give an example to utilize our hybrid waveguide device in practical implementation. We design experimental scheme and simulate the generation of high-dimensional biphoton frequency comb (BFC) through our hybrid waveguide device. Although the hybrid waveguide device is designed for on-chip quantum system, we simulate the experiment in a fiber optic system [39] just as proof-of-concept. The experimental scheme for generating high-dimensional BFC and observing HOM quantum revival is illustrated in Fig. 6. The hybrid waveguide device is pumped by a 780 nm pulse laser with 64 fs pulse duration to generate type-II SPDC photons. A BPF is then placed to block the residual pump light and further eliminate the frequency correlation between signal and idler photons. The SPDC photons can yield purity up to 99.79% with a 10 nm BPF, which lower the experimental requirement for narrow BPF [28]. BFC is generated by passing the SPDC photons through a FFPC, with signal and idler photons in orthogonal polarizations. The FFPC has a free spectral range (FSR) of 15 GHz and bandwidth of 0.5 GHz, respectively. The repetition period T of the BFC will be ~66.7 ps. We assume that signal and idler photons will have the same spectrum after passing through the FFPC without polarization birefringence. Due to the type-II SPDC configuration, there is no probability that both photons will propagate into the same arm of the HOM interferometer. This configuration potentially yields a maximum of 100% visibility of the two-photon interference. The correlation feature in the time-bins of the high-dimensional BFC is characterized by an unbalanced HOM interferometer. The signal and idler photons are sent to different arms of the HOM interferometer. A FPC placed in the lower arm of the interferometer rotates the polarization of idler photons, so that two photons have the same polarization when encountering the 50:50 fiber coupler. An optical delay line is used to tune the relative delay $\Delta T$ between two arms of the HOM interferometer. The coincidences are then collected by single photon counting module with internal electronic delay set to compensate optical delay after 50:50 fiber coupler.

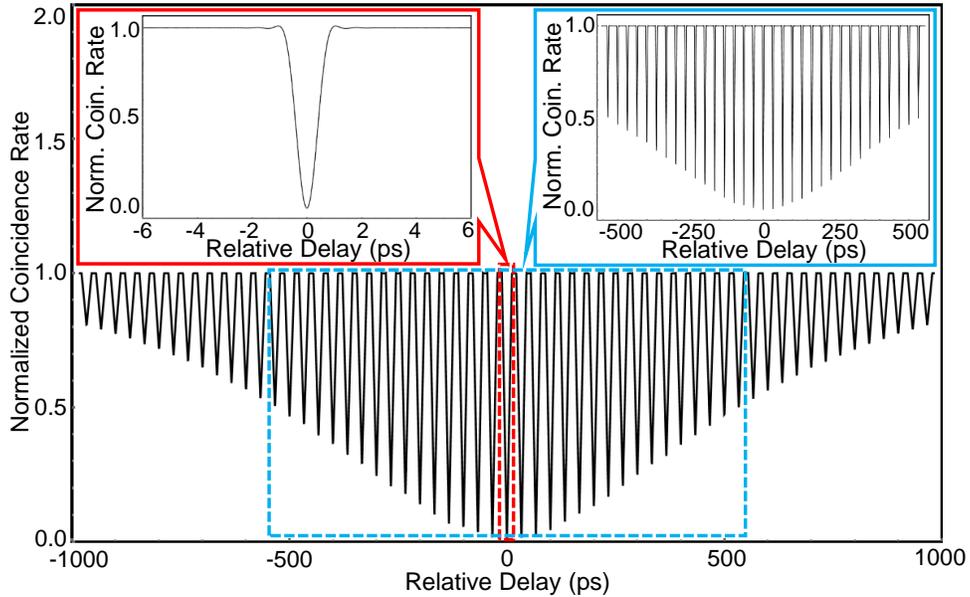

Fig. 7. HOM quantum revival of the high-dimensional BFC. Coincidence rate is computed as a function of relative delay $\Delta T$ between the two arms of the HOM interferometer. Left inset: zoom-in coincidence around zero relative delay between two arms. The base-to-base width of the central dip is estimated to be 1.6 ps. The visibility of the central dip is calculated to be 99.79%. Right inset: zoom-in coincidence for 33 time-bins with interference visibility over 50%.

The simulation results obtained by scanning the optical delay $\Delta T$ from -1000 ps to 1000 ps are shown in Fig. 7. The spacing between dips is estimated to be 33.3 ps, resulting from the FSR of the FFPC [39]. The visibility of the dips decreases exponentially due to the Lorentzian lineshape spectra of the SPDC individual photons after they passing through the FFPC. A zoom-in of the dip around zero delay point is shown in the left inset of Fig. 7. The maximum visibility is calculated to be 99.79%. The base-to-base width of the central dip is estimated to be 1.6 ps, corresponding to the two-photon bandwidth of ~1233 GHz after the BPF. Considering the bin spacing of 15 GHz, 83 frequency-bins can be expected within the phase-matching bandwidth in our experimental scheme. Over the optical delay range from -1000 ps to 1000 ps, 59 dips are observed in simulation, predicting HOM quantum revival at 59 time-bins. Furthermore, as shown in the right inset of Fig. 7, 33 time-bins with visibility beyond classical limit (50%) are obtained, which corresponds to ~5 qubits per photon for high-dimensional entanglement. These high-visibility time-bins enable multi-bit encoding of high-dimensional time-energy entangled photons for entanglement-based quantum key distribution. With proper error correction and privacy amplification, the high-dimensional BFC generated by hybrid waveguide device can achieve higher photon information efficiency and potentially higher key rate for high-dimensional quantum key distribution applications [40, 41].

**Funding**

This work is supported by the National Natural Science Foundation of China (61671090 and 61372037), the Fund of State Key Laboratory of Information Photonics and Optical Communications (Beijing University of Posts and Telecommunications), P. R. China (IPOC2017ZZ04), the National Science Foundation (Emerging Frontiers in Research and Innovation ACQUIRE 1741707 on quantum communications, and awards 1824568, 1810506, 1829071), the University of California National Laboratory research program (LFRP-17-477237), Office of Naval Research (N00014-14-1-0041), and Lawrence Livermore National Laboratory (contract B622827).

**Acknowledgments**

The authors thank Y. Gong, X. Cui and D. Mendinueto for assistance and discussion. Xiang Cheng acknowledges the funding supported by China Scholarship Council.